\documentclass[twocol]{ametsocV5}

\pdfoutput=1 

\usepackage{amsmath,amsfonts,amssymb,bm}
\usepackage{mathptmx}
\usepackage{newtxtext}
\usepackage{newtxmath}
\usepackage[utf8]{inputenc}
\usepackage{color}
\usepackage[squaren]{SIunits}

\newcounter{rupertcommentno}
\newcounter{boualemcommentno}
\newcommand{\thelayer}{{diabatic layer}}
\newcommand{\Thelayer}{{Diabatic layer}}

\newcommand{\dss}{\displaystyle}
\newcommand{\eq}[1]{(\ref{#1})}
\newcommand{\jb}{\!\!}

\newcommand{\syn}{{\rm syn}}

\newcommand{\hor}{\shortparallel}

\newcommand{\rfr}[1]{{#1}_{\rf}^*}
\newcommand{\rf}{{\rm\scriptscriptstyle ref}}

\newcommand{\vect}[1]{{\boldsymbol{#1}}}

\newcommand{\vi}{\vect{i}}
\newcommand{\vj}{\vect{j}}
\newcommand{\vk}{\vect{k}}

\newcommand{\vx}{\vect{x}}

\newcommand{\vp}{\vect{u}}

\newcommand{\myday}{\text{day}}
\newcommand{\hsc}{{h_{\rm sc}^*}}
\newcommand{\hscnostar}{{h_{\rm sc}}}
\newcommand{\Lsyn}{{\ell_{\syn}^*}}

\newcommand{\M}{{\rm M}}
\newcommand{\Ro}{{\rm Ro}}

\newcommand{\eps}{\varepsilon}
\newcommand{\sqeps}{\delta}
\newcommand{\order}[1]{^{(#1)}}
\newcommand{\bigo}[1]{{\cal O}  ( #1 )}
\newcommand{\littleo}[1]{{\scriptstyle \cal O} \big( #1 \big)}
\newcommand{\epsorder}[1]{\eps^{#1}}
\newcommand{\deltaorder}[1]{\sqeps^{#1}}

\newcommand{\zero}{^{(0)}}

\newcommand{\One}{^{(1)}}

\newcommand{\two}{^{(2)}}

\newcommand{\vpprime}{\widetilde{\vp}}

\newcommand{\pinull}{\pi_0}

\newcommand{\pihalf}{\pi_{1/2}}
\newcommand{\pione}{\pi_1}
\newcommand{\pithreehalf}{\pi_{3/2}}

\newcommand{\qv}{{q_v}}
\newcommand{\qs}{\underline{q}}

\newcommand{\cpd}{c_{pd}^*}
\newcommand{\Rd}{R_{d}^*}

\newcommand{\rhonull}{\rho_0}

\newcommand{\rhohalf}{\rho_{1/2}}
\newcommand{\rhoone}{\rho_1}

\newcommand{\Thetabar}{\overline{\Theta}}

\newcommand{\Thetaone}{\Theta_1}

\newcommand{\Thetaprime}{\widetilde{\Theta}}

\newcommand{\dThetaonedz}{{\rm d}\Theta_1/{\rm d}z}
\newcommand{\dzThetaone}{{\Theta_1'}}

\newcommand{\w}{w}

\newcommand{\QTheta}{Q_\Theta}

\newcommand{\diss}[1]{S_{\!\scriptscriptstyle #1}}

\newcommand{\bnabla}{\boldsymbol{\nabla}}

\newcommand{\grad}{{\bnabla}}
\newcommand{\gradhor}{{\bnabla_{\jb\scriptstyle\hor}}}

\newcommand{\dd}[2]{{\frac{d #1}{d #2}}}

\newcommand{\pp}[2]{{\frac{\partial #1}{\partial #2}}}
\newcommand{\pptext}[2]{{\partial #1 / \partial #2}}

\usepackage{color}
\definecolor{white}{gray}{1.0}
\definecolor{light}{gray}{0.50}
\definecolor{verylight}{gray}{0.75}
\definecolor{heavy}{gray}{0.35}
\definecolor{black}{gray}{0.0}
\definecolor{dgreen}{rgb}{0.0,0.7,0}
\definecolor{dred}{rgb}{0.9959,0,0}
\definecolor{green}{rgb}{0.0,0.99599,0.0}
\definecolor{purple}{rgb}{0.6,0.0,0.4}
\definecolor{magenta}{cmyk}{0.1,1.0,0.1,0.1}

\title{QG-DL: Dynamics of a diabatic layer in the quasi-geostrophic framework}

\authors{Rupert Klein\correspondingauthor{Rupert Klein, rupert.klein@math.fu-berlin.de}
and Lisa Schielicke and Stephan Pfahl}

\affiliation{Freie Universität Berlin, Berlin, Germany}

\extraauthor{Boualem Khouider}
\extraaffil{Dept.\ of Mathematics, Univ.\ of Victoria, Victoria, CA}

\abstract{Quasi-geostrophic (QG) theory describes the dynamics of 
synoptic scale flows in the trophosphere that are balanced with respect to both acoustic and 
internal gravity waves. Within this framework, effects of (turbulent) friction near the ground 
are usually represented by Ekman Layer theory. 
The troposphere covers roughly the lowest ten kilometers of the atmosphere while Ekman layer 
heights are typically just a few hundred meters. However, this two-layer asymptotic theory does 
not explicitly account for substantial changes of the potential temperature stratification 
due to diabatic heating associated with cloud formation or with radiative and turbulent heat fluxes, 
which, in the middle latitudes, can be particularly important in about the lowest three 
kilometers. To address this deficiency, this paper extends the classical QG--Ekman layer 
model by introducing an intermediate, dynamically and thermodynamically active layer, called the 
``\thelayer'' (DL) from here on. The flow in this layer is also in acoustic, hydrostatic, and geostrophic 
balance but, in contrast to QG flow, variations of potential temperature are not restricted to small 
deviations from a stable and time independent background stratification. Instead, within the \thelayer, 
diabatic processes are allowed to affect the leading-order stratification. As a consequence, the 
\thelayer\ modifies the pressure field at the top of the Ekman layer, and with it the intensity of 
Ekman pumping seen by the quasi-geostrophic bulk flow. The result is the proposed extended quasi-geostrophic 
three-layer QG-DL-Ekman model for mid-latitude (dry and moist) dynamics.
}

\begin{document}

\maketitle

%
%
%
%
%

%


\section{Introduction}
\label{sec:Intro}


\subsection{Data and Motivation}
\label{ssec:Motivation}

The quasi-geostrophic (QG) theory is one of the most fruitful foundations of theoretical meteorology.   It has guided our understanding of the mid-latitude dynamics of the atmosphere to a large extent and has led to ``potential vorticity (PV) thinking'' as an instructive framework for the interpretation of weather systems \citep{HoskinsEtAl1985}. The standard derivation of QG theory \citep[see, e.g.,][]{Pedlosky1992} captures balanced flow regimes for dry air only, while later extensions include general diabatic source terms and explicit moist process closures \citep[see, e.g.,][and references therein]{SmithStechmann2017}. In recent years there has also been considerable interest in reduced dynamical models that are based on the QG balance and include moist process submodels. Thus \citet{LambaertsEtAl2011,LambaertsEtAl2012,LaineEtAl2011,BembenekEtAl2020} all consider moisture effects on synoptic scales utilizing quasi-geostrophic dynamics, but they all restrict to shallow water, two-, and three-layer models only. A common limitation of such moist extensions of QG theory is that diabatic effects cannot substantially influence the potential temperature background stratification, as further outlined in the sequel. To address this limitation based on systematic asymptotic analysis, the present work proposes a new three-layer asymptotic description (QG-DL-Ekman) involving QG flow in the bulk of the troposphere, an Ekman layer near the ground, and an intermediate dynamically and thermodynamically active \thelayer\ (DL).

To motivate the approach for mid-latitude flows, let us analyze the strength of diabatic effects allowed for in the QG theory. The diabatic source term of interest, $Q_{\Theta}$, appears in the transport equation for the perturbation potential temperature $\widetilde{\Theta}$, 
\begin{equation}
\label{eq:TransportEquationpotentialTemperaturePrelim}
\pp{\widetilde{\Theta}}{t} + \vp\cdot\gradhor{\widetilde{\Theta}} + w\, \pp{\widetilde{\Theta}}{z}
+ w \frac{d\Thetabar}{dz} = Q_{\Theta} + \diss{\Theta}\, ,
\end{equation}
in which $(\vp,w)$ is the vector of horizontal and vertical velocities, $\gradhor$ and $\pptext{}{z}$ are the horizontal gradient and vertical derivative operators, $\Thetabar(z)$ represents the background potential temperature stratification that is independent of time and horizontally homogeneous, and $\diss{\Theta}$ is a general dissipation term, which we do not need to specify further for the present purposes. In the asymptotic limit regime of QG theory, the third term in \eq{eq:TransportEquationpotentialTemperaturePrelim} is negligible whereas the remaining terms on the left are all of the same order of magnitude. Therefore, to assess the characteristic strength of the diabatic heating and dissipation terms on the right as allowed for in QG theory, it suffices to assess the typical magnitude of $\pptext{\widetilde{\Theta}}{t}$ in the QG regime.

The characteristic time scale of quasi-geostrophic motion is given by
\begin{equation}
\rfr{t} = \frac{\Lsyn}{\rfr{u}} =  \frac{\rfr{N}}{f_0^*} \frac{\hsc}{\rfr{u}} = \unit{8 \cdot 10^4}{\second} \sim 1\, \myday, 
\end{equation}
where $\Lsyn = \rfr{N} \hsc / f_0^* = \unit{800}{\kilo\meter}$ is the synoptic length scale, $\rfr{N} = \unit{10^{-2}}{\second^{-1}}$ is a characteristic buoyancy frequency due to the stable background stratification, $f_0^* = \unit{10^{-4}}{\second^{-1}}$ is a typical mid-latitude Coriolis parameter, $\hsc = \unit{8.8}{\kilo\meter}$ is the pressure scale height, and $\rfr{u} = \unit{10}{\meter\per\second}$ is a characteristic flow velocity (see tables~\ref{tab:atmospheric-characteristics} and \ref{tab:reference-values} below, which are adapted from \citep{Klein2010} to match mid-latitude conditions). Now, QG theory results from an asymptotic expansion justified by small Rossby numbers, $\Ro = \frac{\rfr{u}}{f_0^* \Lsyn} = 10^{-1} \ll 1$, and the expansion of potential temperature reads
\begin{equation}\label{eq:QGThetaExpansion}
\Theta = \Theta_0 + \eps \Theta_1(z) + \eps^2\widetilde\Theta(t,\vx, z) + \littleo{\eps^2}\,,
\end{equation}
where $\eps \sim \Ro$ is the small expansion parameter with a constant of proportionality to be specified below. Thus we find that deviations of potential temperature from the stable background distribution are assumed to be $\bigo{\eps^2 \rfr{T}} \sim \unit{3}{\kelvin}$, where $\Theta_0 = \rfr{T} \sim \unit{300}{\kelvin}$ is the reference temperature. These estimates, on the basis of \eq{eq:TransportEquationpotentialTemperaturePrelim}, enable us to assess the characteristic strength of the diabatic source terms allowed for in QG theory as
\begin{equation}\label{eq:SourceStrengthQG}
Q^*_\Theta \sim \frac{\eps^2 \rfr{T}}{\rfr{t}} = \unit{3}{\kelvin\per\myday}\,.
\end{equation}
This level of diabatic heating compares well with the large-scale/daily mean effects of radiation and other physical source terms in reanalysis data \citep{ZhangEtAl2017}. Nevertheless, the physical source terms representing, e.g., latent heat release during cloud formation and turbulent heat fluxes near the surface can be substantially stronger as is also shown in \citep{ZhangEtAl2017}. Further corroboration comes from Figure~4.8 of \citet{Hartmann2016} (based on data from \citet{LettauDavidson1957}) who demonstrates that near-surface temperature over land can vary by more than $\unit{15}{\kelvin}$ between sunrise and early afternoon, which implies heating rates of more than $\unit{45}{\kelvin\per\myday}$ near the surface, and still more than $\unit{12}{\kelvin\per\myday}$ above $\unit{500}{\meter}$ altitude. Finally, radiative heating or cooling rates can largely exceed the $\unit{3}{\kelvin\per\myday}$ average, in particular, near the upper and lower edges of clouds. For example, \citet[][see their Fig.~8]{MatherEtAl2007}, based on remote sensing observations, estimate heating rates in excess of $\unit{20}{\kelvin\per\myday}$ near the base and cooling rates of similar magnitudes near the top of clouds in the tropical upper troposphere.

These considerations and estimates imply, that 
\begin{enumerate}

\item QG theory describes only higher order (i.e., small) deviations of potential temperature from a given, time independent background stratification, and

\item the intensity of diabatic processes allowed for in QG theory amount to potential temperature tendencies of $\sim \unit{3}{\kelvin\per\myday}$. However, heating rates associated with surface-atmosphere fluxes or radiation near cloud edges can be much larger. 

\end{enumerate}
Therefore, QG scaling is inadequate to describe weather with substantially stronger influence of diabatic processes.

These two implications are not restricted to the classical dry-air QG system. Recent extensions of the QG theory to moist dynamics by \citet{SmithStechmann2017,MarsicoEtAl2019} are subject to the same limitations. Thus, in their section~3, \citet{SmithStechmann2017} state that ``appropriately defined Rossby and Froude numbers are comparable and small'', and that one key assumption of their theory is that the background stratification of the equivalent potential temperature stratification is large. As a consequence, with respect to the transport of buoyancy their theory follows closely the lines of classical QG theory. The immediate implication is again that, within their model, the dynamics cannot substantially change the stability of the stratification.

Let us support our point regarding the strength of moisture effects in the lowest few kilometers of the atmosphere by two additational estimates. First, we assess a mean precipitation rate that would correspond to the typical QG-scale heating rate of approximately $\unit{3}{\kelvin\per \myday}$ when generated by latent heat release and the subsequent fall-out of rain. With $\rfr{L} = \unit{2.5 \cdot 10^6}{\joule\per\kilo\gram}$ the latent heat of condensation of water vapor, $\cpd \sim \unit{10^3}{\joule\per\kilo\gram\per\kelvin}$ the dry air heat capacity at constant pressure, $\rho_d^* \sim \unit{1.25}{\kilo\gram\per\meter^3}$ and $\rho_{l}^* \sim \unit{10^3}{\kilo\gram\per\meter^3}$ the typical dry air and liquid water densities, respectively, and $A$ some horizontal reference area on the synoptic scale, we let 
\begin{equation}
\rho_d^*\, \cpd \, Q^*_\Theta \, \hsc \, A = \rho_{l}^*\, \rfr{L}\, P_{\text{QG}} \, A\, .
\end{equation}
Here, the left hand side is the rate of change of dry air internal energy within a volume of $\hsc A$ due to the diabatic heating $Q^*_\Theta$, while the right hand side represents the rate of latent heat released by the condensation of water forming precipitation. 
The rate of precipitation, $P_{\text{QG}}$, is measured in $\milli\meter\per\hour$. 
This yields, given the previous concrete data with a diabatic heating of $Q_\Theta^* = \unit{3}{\kelvin\per \myday}$, a characteristic precipitation rate under QG scaling of 
\begin{equation}\label{eq:QGPrecipitationStrength}
P_{\text{QG}}
= \frac{\rho_d^*}{\rho_{l}^*} \, \frac{\cpd \, Q^*_\Theta}{\rfr{L}}  \, \hsc  
\sim  \unit{0.4}{\milli\meter\per\hour}\, .
\end{equation}
This is a very small number given that even the category of ``light rain'' is associated with precipitation rates of $\unit{2.5}{\milli\meter\per\hour}$ \citep{AMSGlossary}. 

\begin{figure}[htbp]
\begin{center}
\includegraphics[width=0.4\textwidth]{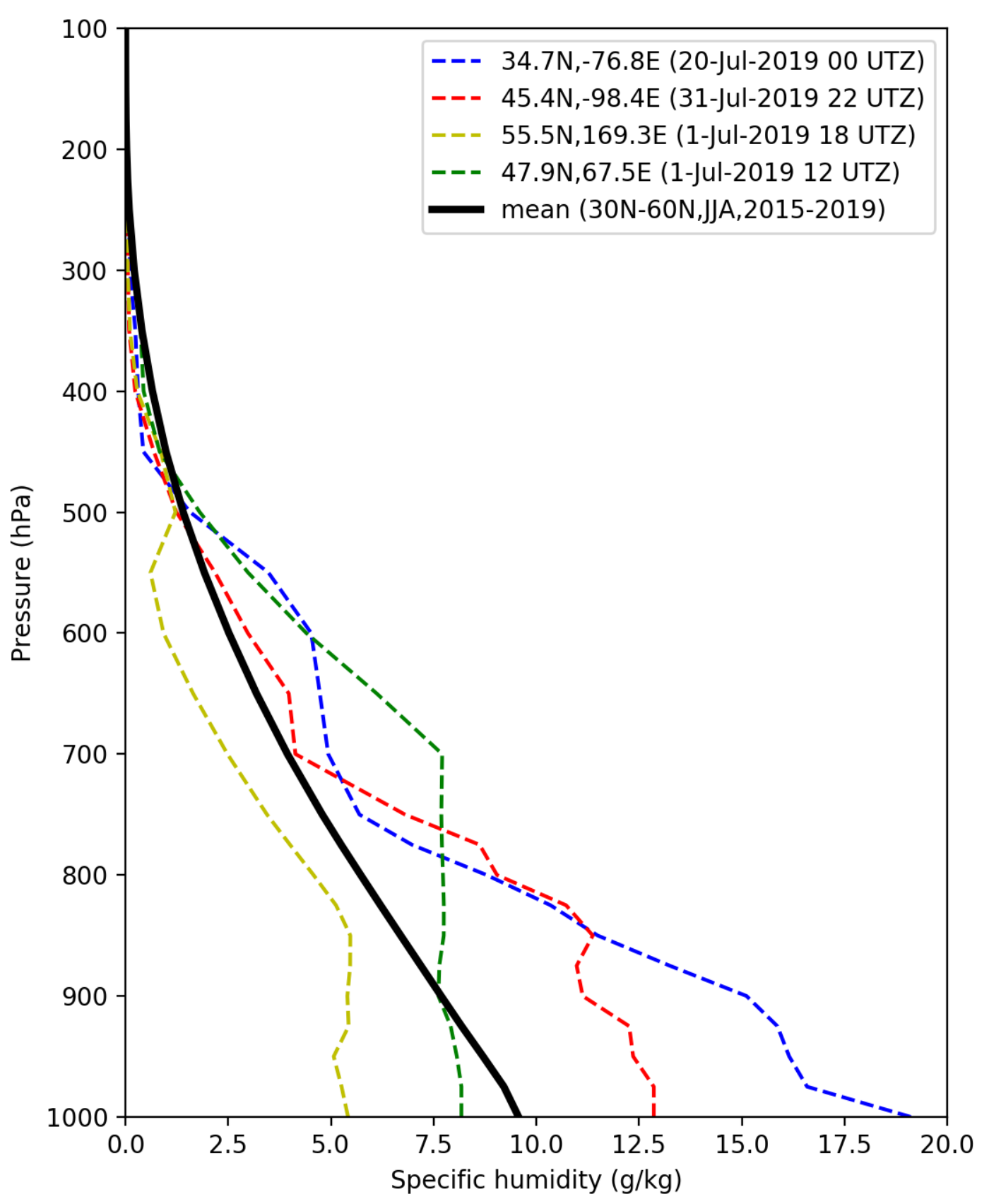}
\caption{Vertical profiles of specific humidity derived from the global ERA5 Reanalysis data set \citep{HersbachEtAl2020} for specific locations in the midlatitudes at some selected times during July 2019 (dashed lines).  The solid black line displays the mean vertical profile of the boreal summer months June to August in the 5-year period 2015-2019 calculated for latitudes between 30$^{\circ}$N and 60$^{\circ}$N.}
\label{fig:MoistureProfiles}
\end{center}
\end{figure}

Secondly, we observe that in many mid-latitude situations most of the atmospheric water content resides in the lowest few kilometers while it falls off rapidly with height. See fig.~\ref{fig:MoistureProfiles} for some instantaneous and time-averaged vertical profiles of specific humidity in the mid-latitudes extracted from the ERA5 reanalysis data set \citep{HersbachEtAl2020}. Although individual profiles more or less differ in terms of magnitude and distribution of specific humidity in the lower troposphere, in most of the profiles shown specific humidity is down to a quarter of its maximum value at heights of $700$ to $\unit{600}{\hecto\pascal}$, i.e., at about $3$ to $\unit{4}{\kilo\meter}$.

What is the precipitation rate that would be generated if a substantial fraction of the water vapor with a mixing ratio at the level of $q^*_{v} \sim \unit{10}{\frac{\gram}{\kilo\gram}} = 0.01$ resides in the lowest, say three, kilometers ($\sim \sqrt{\eps}\hsc$) of the atmosphere initially (i.e., in the \thelayer) and if some frontal or related process converts it into precipitation over the time scale of a day? The corresponding rain mass flux density, ${\dot M}''$, would amount to
\begin{equation}
{\dot M}''
=
\frac{\rho^*_d q^*_{v}}{\Lsyn/\rfr{u}} \sqrt{\eps} \hsc 
\sim 
\unit{\frac{1}{15} \cdot 10^{-2}}{\frac{\kilo\gram}{\meter^2 \second}}\,,
\end{equation}
which translates to a precipitation rate of
\begin{equation}
P_{\text{DL}} = \frac{{\dot M}''}{\rho^*_{l}} \cdot 3600\, \second\per\hour  \cdot 1000 \, \milli\meter\per\meter
\sim \unit{2.4}{\milli\meter\per\hour}\,.
\end{equation}
This is substantially larger than the QG estimate in \eq{eq:QGPrecipitationStrength}, and, e.g., in line with typical mean precipitation rates within midlatitude cyclones as shown, for instance, by \citet{BengtssonEtAl2009} who obtained averaged precipitation intensities in the range of $\unit{1.0...3.6}{\milli\meter\per\hour}$ in a composite analysis based on climate simulations and ERA40 reanalysis data. 

Much of the enhanced diabatic heating described above, except for the radiative heating at cloud edges, is restricted mostly to the lower part of the troposphere. On the one hand, this is related to the fact that the atmospheric moisture content is a strong function of temperature and thus decreases with altitude (see again fig.~\ref{fig:MoistureProfiles}), thereby constraining the latent heat that can be released at higher levels. On the other hand, turbulent convective heat fluxes are strongest near the surface and generally decrease in magnitude with height. In particular, turbulent fluxes over land are associated with the formation of mixing layers with typical heights of $\unit{1.5...2.5}{\kilo\meter}$ \citep{WangWang2014}. These layers are characterized by approximately constant values of potential temperature and specific humidity with height, while the humidity drops rapidly above the mixing layer top. For the incorporation of such strong diabatic heating in existing QG theories, it thus appears reasonable to focus on a ``diabatic layer'' with a vertical extent of $\sim \unit{3}{\kilo\meter}$ above the surface.


\subsection{Organization of the paper}
\label{ssec:Orga}

The rest of the paper is organized as follows. Section~\ref{sec:Essence} describes and summarizes the essential properties of the diabatic layer. Section~\ref{sec:Methodology} provides the governing equations and asymptotic scalings adopted in the sequel. Section~\ref{sec:QGAndEkman} contains a brief overview of the classical QG--Ekman layer theory, which we include here for the sake of completeness and to render the paper largely self-contained. Section~\ref{sec:WeatherLayer} derives the \thelayer\ equations.
Section~\ref{sec:SummaryAndDiscussion} provides further discussion, a summary, and an outlook. 


\section{The essence of the \thelayer\ (DL) dynamics}
\label{sec:Essence}

Here we summarize the essential differences between the new QG-DL-Ekman and the classical QG-Ekman models. Equations in this section are in dimensional form, but we have dropped the $^*$-indicator here to streamline the notation. 

\begin{figure}[htbp]
\begin{center}
\includegraphics[width=0.5\textwidth]{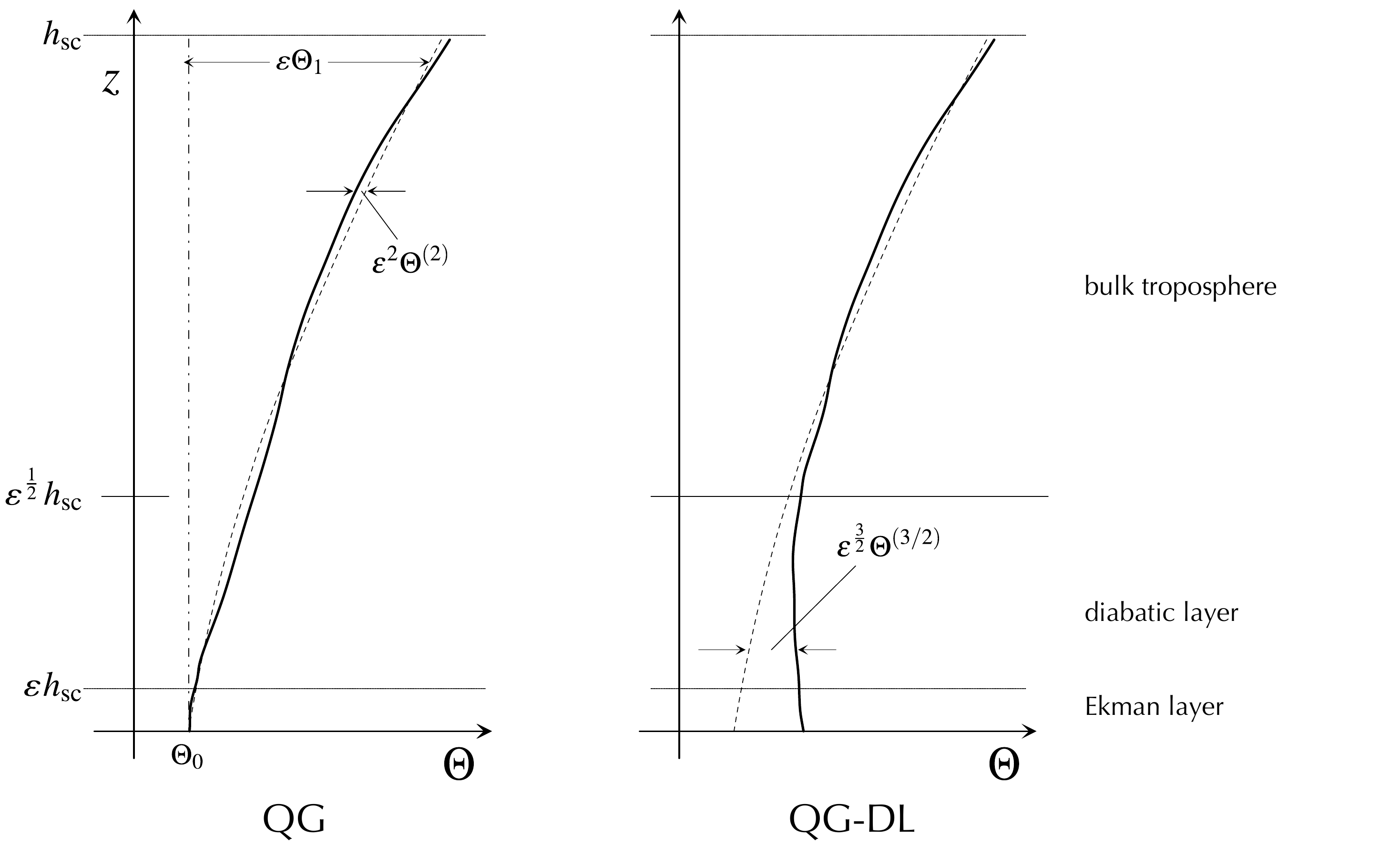}
\caption{Magnitudes of the dynamically evolving deviations of potential temperature from the background stratification $\Theta_0 + \eps\Theta_1(z)$ (dashed line) across the pressure scale height according to the QG (left) and the present three-layer theory (right). Whereas QG theory always assumes a stable background stratification, $d\Theta_1/dz > 0$, dynamic variations of potential temperature in the \thelayer\ are comparable to the variability of the background stratification, so that it can accomodate neutral or even unstable stratification.}
\label{fig:QG_vs_QGWL}
\end{center}
\end{figure}

Classical QG theory divides the trophosphere into two dynamically different vertical layers \citep{Pedlosky1992}. In the bulk of the troposphere, friction and turbulent transport are neglected to leading order and the horizontal momentum balance is dominated by the pressure gradient and Coriolis terms. Near the ground, these terms are important, however, and enter into a three-term balance with the former two in the ``Ekman layer''. As indicated above, the scalings underlying that theory only allow for rather weak diabatic effects. In particular, these are not strong enough to substantially change the dominant potential temperature stratification, which is represented by the terms $\Theta_0 + \eps\Theta_1(z)$ in \eq{eq:QGThetaExpansion}. In fact, space-time dependent variability of the potential temperature arises first at $\bigo{\eps^2}$. The situation is sketched in the left panel of fig.~\ref{fig:QG_vs_QGWL}. Yet, in the lower few kilometers, thermal or moist convection can cumulatively change the mean stratification all the way to nearly neutral \citep{Stevens2005} on (sub-)daily time scales and this is not covered by QG theory.  

A remedy is proposed here by introduction of an additional ``\thelayer''. As shown on the right panel of fig.~\ref{fig:QG_vs_QGWL}, we include a layer of an intermediate height of $\sqrt{\eps} \hscnostar \sim \unit{3}{\kilo\meter}$. Within this layer all deviations from the reference potential temperature $\Theta_0$ are subject to the layer's dynamics, and not just higher-order deviations from the leading and first order background stratification $\Theta_0 + \eps\Theta_1(z)$ as is the case in QG theory, see \eq{eq:QGThetaExpansion}. Mathematically, this change in scalings is represented by a modified expansion scheme, $\Theta = \Theta_0 + \eps^{\frac{3}{2}}\widetilde{\Theta}(t,\vx,\eta) + \text{h.o.t.}$ ($\text{h.o.t.} = \text{higher order terms}$) for the potential temperature in the \thelayer, with the scaled vertical coordinate $\eta = z/\sqrt{\eps}$. With this ansatz, the potential temperature transport equation becomes 
\begin{equation}
\label{eq:TransportEquationpotentialTemperatureQGWL}
\pp{\widetilde{\Theta}}{t} + \vp\cdot\gradhor{\widetilde{\Theta}} = Q_{\Theta} + \diss{\Theta}\, ,
\end{equation}
from which the vertical advection terms are absent. Clearly, in the absence of the source and dissipation terms on the right, and for horizontally homogeneous initial data $\widetilde{\Theta}(0,\vx,\eta) = \widetilde{\Theta}_*(\eta)$, $\widetilde{\Theta}$ will remain constant in time and will just maintain the time independent background stratification, which is compatible with QG theory. The theory does allow for non-zero diabatic source terms, however, and this generates the leading-order variability of the potential temperature soundings which QG theory cannot account for (cf.\ the right panel of fig.~\ref{fig:QG_vs_QGWL}).

The horizontal wind in the \thelayer\ is determined by geostrophic balance, i.e., by
\begin{equation}\label{eq:GeostrophicBalanceQGWL}
f_0 \, \vk \times \vp + c_p \Theta_0 \gradhor{\pi} = 0\, ,
\end{equation}
and the (Exner) pressure perturbation, $\pi$, is subject to hydrostatic balance, i.e., 
\begin{equation}\label{eq:HydrostaticBalanceQGWL}
c_{p} \Theta_0 \pp{\pi}{\eta} = g\frac{\widetilde{\Theta}}{\Theta_0}\,.
\end{equation}
Both these balances are also found in QG theory. The next essential difference between the two regimes is that, the \thelayer\ being a boundary layer in the sense of matched asymptotic expansions, the pressure near the top of the layer has to agree asymptotically with that near the bottom of the QG bulk flow. Therefore, the \thelayer\ pressure is determined by the pressure at the bottom of the QG layer together with the hydrostatic relation in \eq{eq:HydrostaticBalanceQGWL}. This is in contrast to QG theory in which the pressure is determined as the solution to a three-dimensional Poisson-type equation with the potential vorticity as a source term \citep{Pedlosky1992}. 

Due to the hydrostatic balance in \eq{eq:HydrostaticBalanceQGWL}, the pressure at the bottom of the \thelayer\ will generally differ substantially from that found at the bottom of the QG layer. But it is the former which drives the Ekman layer, so that the forcing by the \thelayer\ will generate a horizontal distribution of Ekman pumping that deviates strongly from the pumping that would be induced by the QG pressure field. 

\begin{figure}[htbp]
\begin{center}
\includegraphics[width=0.5\textwidth]{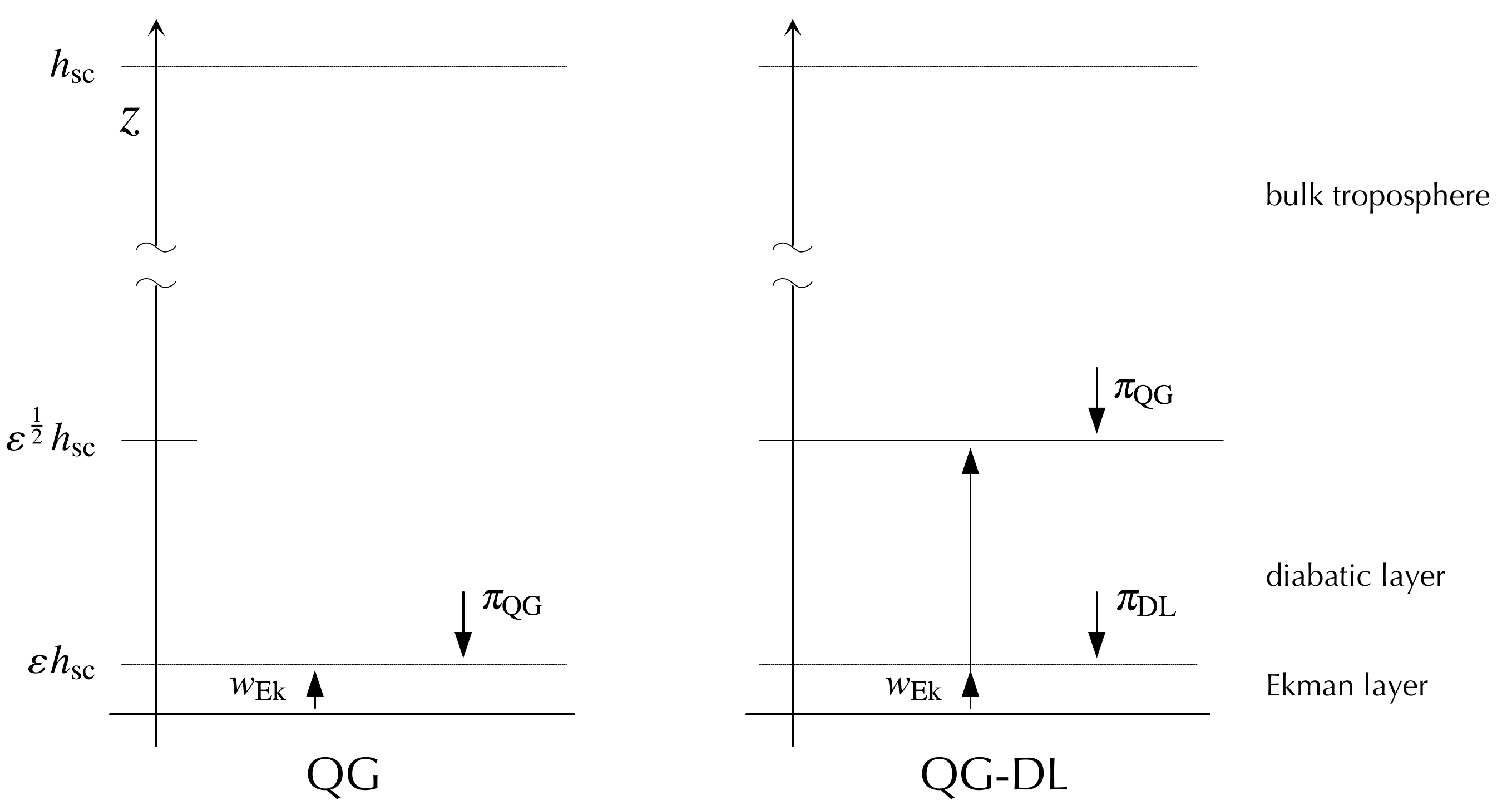}
\caption{Feedbacks between the QG flow in the bulk of the trophosphere and the underlying boundary layer(s) for QG (left) and QG-DL (right).}
\label{fig:Feedbacks}
\end{center}
\end{figure}

Next it turns out, as we show below, that the leading \emph{and} first order horizontal flow fields in the \thelayer\ are divergence-free. As a consequence, the leading order vertical velocity is constrained by mass conservation to be constant throughout the \thelayer, i.e., 
\begin{equation}
\pp{w}{\eta} = 0\,.
\end{equation}
This implies that the vertical velocity at the top of the Ekman layer is maintained through the \thelayer\ and provides the bottom boundary condition for the vertical velocity in the QG domain. In short, the \thelayer\ substantially influences Ekman pumping, and hence the QG flow aloft. The latter, in turn, controls the pressure field at the top of the \thelayer, thereby closing a feedback loop of couplings between the layers. Figure~\ref{fig:Feedbacks} summarizes these causal feedbacks and compares them with those found in QG theory. 
%
  

\section{Dimensionless governing equations and distinguished limits} 
\label{sec:Methodology}

The dimensionless inviscid rotating compressible flow equations in the beta plane approximation, including general diabatic source and transport terms in the thermodynamic equation, i.e.,
\begin{subequations}\label{eq:NondimensionalCompressibleFlowEquations}
  \begin{alignat}{6}
    \label{eq:HorizontalMomentumEquation}
    &\dd{\vp}{t} 
      &
        & 
          &
            & 
              & 
                & + \frac{1}{\eps^3} \, \frac{\Theta}{\Gamma} \, \gradhor{\pi} 
                  &
                    & + \frac{1}{\eps} \,
                        f \, \vect{k} \times \vp
                      & 
                        & = \diss{\vp}  \, , 
                          \\
    \label{eq:VerticalMomentumEquation}
    &\dd{w}{t}
      &
        & 
          &
            & 
              &
                & + \frac{1}{\eps^5}\,  \frac{\Theta}{\Gamma}\, \pp{\pi}{z} 
                 &
                   &  
                     &
                       & = - \frac{1}{\eps^5} + \diss{w}\, , 
                         \\ 
    \label{eq:ContinuityEquation}
    &\dd{\rho}{t}
      &
        & 
          &
            & 
              &
                & + \rho\,  \gradhor\cdot \vp  
                  &
                    & + \rho \pp{w}{z}
                      & 
                        & = 0\, ,
                          \\
    \label{eq:TransportEquationpotentialTemperature}
    &\dd{\widetilde{\Theta}}{t}
      &
        & 
          &
            & 
              &
                & + w \frac{d\Theta_1}{dz}
                  &
                    &
                      &
                        & = Q_{\Theta} + \diss{\Theta}\, ,
                          \\
    \label{eq:EquationOfState}
    &
      &
        &
          &
            &
              &
                &
                  &
                    & \eps\left( \widetilde\Theta + \Theta_1\right) 
                      & 
                        & = \frac{\pi^{\frac{1}{\gamma-1}}}{\rho} - 1\, .
  \end{alignat}
\end{subequations}
are our point of departure, with $\eps$ our main dimensionless parameter which sets the limit regime for the asymptotic expansions and is further explained in \eq{eq:distinguished_limit} below, and
\begin{equation}
\dd{}{t} = \pp{}{t} + \vp\cdot\gradhor + w \pp{}{z}\,.
\end{equation}
The physical meaning of all variables is summarized in table~\ref{tab:Variables}. These equations follow \citep{PaeschkeEtAl2012}, except that we have introduced the Exner pressure $\pi$ instead of the thermodynamic pressure $p$, see \eq{eq:dimensionless_dependent_variables} below. 

In \eq{eq:NondimensionalCompressibleFlowEquations},
\begin{equation}\label{eq:partial_derivatives}
\begin{array}{c}
\dss
\gradhor{} = \vi\, \partial/\partial x + \vj\, \partial/\partial y\, , 
\quad
\grad = \gradhor{} + \vk\,\partial/\partial z\, , 
\end{array}
\end{equation}
denote the horizontal and full three-dimensional gradients, respectively. 

Note that, although the present paper is partially motivated by the aim of including thermal effects of moist processes at realistic magnitudes in a balanced flow model, we have omitted the moisture transport equations and moist thermodynamic equations of state in \eq{eq:NondimensionalCompressibleFlowEquations} to avoid unnecessarily lenghty formal developments. In fact, our main aim is to demonstrate how strong diabatic effects affecting the lower few kilometers of the atmosphere can be included in a QG framework in general, whereas the specifics of moist processes are not essential to our arguments.

Thus, letting asteriscs denote dimensional quantities, the variables 
\begin{equation}\label{eq:dimensionless_dependent_variables}
\begin{array}{c}
\dss
\pi = \left(\frac{p^*}{\rfr{p}}\right)^{\Gamma}\, ,
\quad
\rho = \frac{\rho^*}{\rfr{\rho}}\, ,
\quad
\Theta = \frac{\Theta^*}{\rfr{T}}\, ,
  \\
\dss \vp = \frac{\vp^*}{\rfr{u}}\, ,
\quad
w = \frac{w^*}{\rfr{u}} \frac{\rfr{\ell}}{\hsc}\, , 
\end{array}
\end{equation}
in \eq{eq:NondimensionalCompressibleFlowEquations} are the dimensionless Exner pressure, density, dry air potential temperature, and the horizontal and vertical velocities, respectively, while
\begin{equation}\label{eq:GammaDefinition}
\Gamma = \frac{\Rd}{\cpd} \equiv \frac{\gamma-1}{\gamma}
\end{equation}
is the ratio of the dry air heat capacity at constant pressure and dry air gas constant, where $\gamma$ is the dry air isentropic exponent. The dimensionless time and the horizontal and vertical coordinates used in 
\eq{eq:NondimensionalCompressibleFlowEquations} and the subsequent equations are 
\begin{equation}\label{eq:dimensionless_independent_variables}
\begin{array}{c}
\dss
t = \frac{t^*\rfr{u}}{\rfr{\ell}}\, , 
\quad
\vx = (x, y) = \frac{\vx^*}{\rfr{\ell}}\, , 
\quad
z = \frac{z^*}{\hsc}\,.
\end{array}
\end{equation}

To obtain the exact form of \eq{eq:NondimensionalCompressibleFlowEquations} in which $\eps \ll 1$ appears as the sole dimensionless small parameter giving rise to asymptotic expansions, the following distinguished limits for the Mach, Froude, and Rossby numbers have been adopted:
\begin{equation}\label{eq:distinguished_limit}
\begin{array}{c}
\dss
\M^2 = \frac{\rfr{\rho}{\rfr{\vp}}^2}{\rfr{p}} = \eps^3\,, 
\quad
\Ro = \frac{\rfr{u}}{f_0^* \rfr{\ell}} =  \frac{\eps}{f_0}\, , 
  \\
\dss 
\frac{\beta^* {\rfr{\ell}}^2}{\rfr{u}} = \eps^2 \beta \, . 
\quad
\frac{\hsc}{\rfr{\ell}} = \eps^2\, .
\end{array}
\end{equation}
Here, $f_0$ and $\beta$ are $\bigo{1}$ as $\eps\to 0$. Furthermore, \eq{eq:EquationOfState}  encodes the assumption that deviations of the dimensionless potential temperature from unity are $\Delta \Theta^*/\rfr{T} = \bigo{\eps}$. Via the scaling of the  buoyancy frequency $\rfr{N}$, this provides an interpretation of the chosen horizontal scale $\rfr{\ell}$ as follows: Since
\begin{equation}\label{eq:WeakStratification}
\rfr{N} \sim \sqrt{\frac{g^*}{\rfr{T}}\frac{\Delta \Theta^*}{\hsc}}
\sim \sqrt{ \eps \frac{g^*}{\hsc}}\,,
\end{equation}
we find, utilizing \eq{eq:distinguished_limit} and \eq{eq:WeakStratification}, 
\begin{equation}
\frac{\rfr{N} \hsc}{f_0^* \rfr{\ell}} \sim \sqrt{\eps} \frac{\sqrt{g^*\hsc}}{f_0^* \rfr{\ell}} 
\sim \sqrt{\eps} \frac{\Ro}{\M} = \frac{1}{f_0} = \bigo{1} \ \ (\eps \to 0)\,.
\end{equation}
That is, $\rfr{\ell} = \Lsyn/f_0$, where $\Lsyn = \rfr{N} \hsc / f_0^*$ is the usual synoptic length scale and $f_0$ is the $\eps$-independent dimensionless scaling factor first mentioned in the context of \eq{eq:QGThetaExpansion} above. As a consequence, the reference length and time scales chosen here are compatible with those adopted by \citet{Pedlosky1992} in his textbook derivation of the quasi-geostrophic theory.

Table \ref{tab:atmospheric-characteristics} lists the general characteristics of the mid-latitude atmosphere (latitude $\phi = 45\degree N$) which we have combined in Table \ref{tab:reference-values} to obtain the reference values for non-dimensionalization. In particular, $\rfr{u}$ is an estimate of the thermal wind shear due to the equator-to-pole potential temperature difference \citep{Klein2010}. The latter happens to coincide in magnitude with both the vertical potential temperature variation across the troposphere \citep[see, e.g.,][]{HeldSuarez1994}, and with temperature changes associated with the latent heat of moisture, $\Delta\Theta^* \sim \rfr{L}\rfr{\qv}/\cpd$. Here, $\rfr{\qv} \sim \unit{10^{-2}}{\kilo\gram\per\kilo\gram}$ is a typical saturation water vapor mixing ratio in the mid-latitudes. Quantitatively, this amounts to 
\begin{equation}
\label{eq:EpsMagnitude}
\eps \sim \frac{\Delta\Theta^*}{\rfr{T}} \sim \frac{1}{10}
\end{equation}
for the chosen atmospheric reference conditions. This corresponds to a distinguished limit tying latent heating, background stratification, and Rossby number to the expansion parameter $\eps$ via 
\begin{equation}
\frac{\Delta\Theta^*}{\rfr{T}} \sim \frac{\hsc{\rfr{N}}^2}{g^*} \sim \Ro \sim \eps\,. 
\end{equation}
See also \citep{Klein2010,SmithStechmann2017,HittmeirKlein2018}.

\begin{table}
\begin{center}
\begin{tabular}{l@{\quad}l}
\text{\bfseries dependent variables}
  \span\omit
    \\
$\vp$
  & horizontal velocity vector
    \\
$w$
  & vertical velocity
    \\
$\rho$
  & density
    \\
$\Theta$
  & potential temperature
    \\
$\Theta_1$
  & background stratification of potential temperature 
    \\
  & (first order in $\eps$)
    \\
$\widetilde{\Theta}$
  & perturbation potential temperature (first order in $\eps$)
    \\
$\pi$
  & Exner pressure
    \\
$\qs$
  & $= (q_v, q_c, q_r)$ \ vapor, cloud, and rain water mixing ratios 
    \\
$S_{\xi}$
  & turbulent transport term for variable $\xi$
    \\
$Q_{\xi}$
  & source term for variable $\xi$
    \\[3pt]
\text{\bfseries independent variables}
  \span\omit
    \\
$t$
  & time
    \\
z & vertical space coordinate
    \\
$\vx$
  & $= (x, y)$ \ horizontal coordinates 
\end{tabular}
\end{center}
\caption{Physical meaning of the variables in \eq{eq:NondimensionalCompressibleFlowEquations}\label{tab:Variables}}
\end{table}%
\begin{table}
  \begin{center}
    \begin{tabular}{@{}l@{\quad }l@{\ }c@{\ }r@{\ }l@{\ }r@{\ \ }l}
      Gravitational acceleration                       
        & $g^*$ 
          & $=$ 
            &  
              & 
                & $9.81$  
                  & \meter\,\second$^{-2}$
                    \\[0pt]
      Coriolis parameter ($\phi = 45\degree$ N)                      
        & $f_0^*$
          & $=$ 
            &  
              &  
                & $10^{-4}$
                  & \second$^{-1}$
                    \\[0pt]
       $(df^*/dy)_0$ ($\phi = 45\degree$ N)                    
        & $\beta^*$
          & $=$ 
            &   
              & 
                & $8.0\cdot 10^{-12}$
                  & \meter$^{-1}$\,\second$^{-1}$
                    \\[0pt]
      Pressure                     
        & $\rfr{p}$    
          & $=$ 
            &  
              & 
                & $10^5$
                  & \pascal
                    \\[0pt]
      Temperature                  
        & $\rfr{T}$    
          & $=$ 
            & 
              &  
                & $300$
                  & \kelvin
                    \\[0pt]
      Brunt-V\"ais\"al\"a frequency                  
        & $\rfr{N}$    
          & $=$ 
            & 
              &  
                & $10^{-2}$
                  & \second$^{-1}$
                    \\[0pt]
      Dry gas constant             
        & $\Rd$
          & $=$ 
            & 
              & 
                & $287$
                  & \meter\,\rpsquare\second\,\kelvin$^{-1}$
                    \\[0pt]
      Latent heat of condensation                     
        & $\rfr{L}$ 
          & $=$ 
            &  
              &   
                & $2.5$ 
                  & \joule\per\kilo\gram
                    \\[0pt]
      water component mixing ratio                     
        & $\rfr{q}$ 
          & $=$ 
            &  
              &   
                & $10$ 
                  & \gram\per\kilo\gram
                    \\[0pt]
      Isentropic exponent                      
        & $\gamma$ 
          & $=$ 
            &  
              &   
                & $1.4$ 
                  & 
                    \\[0pt]
      \end{tabular}
    \end{center}
    \caption{Characteristic atmospheric flow parameters\label{tab:atmospheric-characteristics}}
\end{table}
\begin{table}
  \begin{center}
    \begin{tabular}{@{}l@{\ \ \ }l@{\ }c@{\ }r@{\ }l@{\ \ }r@{\ }l}
      Density                      
        & $\rfr{\rho}$ 
          & $=$ 
            & $\dss \frac{\rfr{p}}{R\rfr{T}}$
              & $\sim$ 
                & $1.16$ 
                  & \kilo\gram\,\meter$^{-3}$ 
                    \\[0pt]
      Horizontal velocity                   
        & $\rfr{u}$ 
          & $=$ 
            & $\dss \frac{\tan\phi}{\pi/2}\frac{N^2}{f_0^2}\beta\hsc^2$
              & $\sim$
                & $10$
                  & \meter\,\second$^{-1}$
                    \\[0pt]
      Vertical velocity                     
        & $\rfr{w}$ 
          & $=$ 
            & $\dss \frac{\hsc}{\Lsyn}\ \rfr{u}$
              & $\sim$
                & $0.05$
                  & \meter\,\second$^{-1}$
                    \\[0pt]
      Horizontal distance                       
        & $\Lsyn$ 
          & $=$ 
            & $\dss \frac{N}{f_0} \hsc$ 
              & $\sim$
                & $800$  
                  & \kilo\meter
                    \\[0pt]
      Vertical distance                       
        & $\hsc$ 
          & $=$ 
            & $\dss \frac{\rfr{p}}{g\rfr{\rho}}$ 
              & $\sim$
                & $8.1$  
                  & \kilo\meter
                    \\[0pt]
      Time                         
        & $\rfr{t}$
          & $=$ 
            & $\dss \frac{\Lsyn}{\rfr{u}}$
              & $\sim$
                & $8 \cdot 10^4$
                  & \second
                    \\
      \end{tabular}
    \end{center}
    \caption{Further derived reference values yielding \eq{eq:NondimensionalCompressibleFlowEquations}\label{tab:reference-values}}
\end{table}

We remark that the atmospheric flow parameters in table~\ref{tab:atmospheric-characteristics} are essentially equivalent to those used in the general modelling framework in \citep{Klein2010}, while the reference quantities for nondimensionalization in table \ref{tab:reference-values} have been constructed from these to fit the present application and streamline the subsequent developments.


\section{The Quasi-geostrophic (QG) and Ekman Layer theories}
\label{sec:QGAndEkman}

\subsection{Quasi-geostrophic (QG) flow}
\label{ssec:QG}

Here we rederive the QG theory following \citep{Pedlosky1992} neglecting diabatic effects 
to highlight the essence of the argument. The analysis proceeds with the expansion scheme 
\begin{subequations}\label{eq:SynopticExpansions}
\begin{alignat}{8}
  \label{Syn:PressureExpansion}
  &\pi  
  && = \pinull
  && + \varepsilon\pione 
  && + \epsorder{2}\, \left(\pi\order{2} + \pi_2\right)
  && + \littleo{\epsorder{2}}
  \\
  \label{Syn:DensityExpansion}
  & \rho 
  && = \rhonull
  && + \varepsilon\rhoone
  && + \epsorder{2}\, \left( \rho\order{2} + \rho_2\right)
  && + \littleo{\epsorder{2}}
  \\
  \label{Syn:ThetaExpansion}
  & \Theta
  && = 1 
  && + \varepsilon \Thetaone
  && + \epsorder{2}\, \left( \Theta\order{2}+ \Theta_2\right)
  && + \littleo{\epsorder{2}}
\end{alignat}
\begin{alignat}{4}
\label{Syn:HorizontalVelocity}
  & \vp
  && = \phantom{\epsorder{3}}\vp\zero
  && + \varepsilon\vp\One
  && + \littleo{\eps}
  \\
\label{Syn:VerticalVelocity}
  & w
  && = 
  && \phantom{+ \ }  \eps w\order{1} 
  && + \littleo{\eps}   
\end{alignat}
\end{subequations}
where $(\pi_{i},\rho_{i},\Theta_{i})(z)$ and $\left(\vp\order{i}, w\order{i}, \pi\order{i}, \rho\order{i}, \Theta\order{i}\right)(t,\vx,z)$ represent the mean background state and deviations from it, respectively.  Note that we work with dimensionless variables here, and that the potential temperature is non-dimensionalized by $\rfr{T}$. Since the potential temperature is also constant in the vertical at leading order, the leading term in \eq{Syn:ThetaExpansion} is just unity. Due to the effects of gravity, however, the leading order Exner pressure $\pi_0(z)$ and density $\rho_0(z)$ do exhibit vertical variations at leading order. For these variables, the non-dimensionalization merely amounts to $\pi_0(0) = \rho_0(0) = 1$ at the surface.

From the scalings of the pressure gradient and Coriolis terms in the horizontal momentum balance in \eq{eq:HorizontalMomentumEquation} it follows that, under leading-order geostrophic balance, deviations of the pressure from the background state will arise first at order $\bigo{\eps^2}$. (The Coriolis term is $\bigo{1/\eps}$, yet the pressure gradient term has a prefactor $1/\eps^3$. Thus, for these two terms to  balance each other, pressure variations have to be of $\bigo{\eps^2}$.) The vertical momentum equation in \eq{eq:VerticalMomentumEquation} is dominated by hydrostatic balance up to at least five orders in $\eps$ and, as a consequence, deviations of density from the background state follow the pressure scaling and also start at $\bigo{\eps^2}$. This explains the expansion schemes in \eq{Syn:PressureExpansion}--\eq{Syn:ThetaExpansion}. In \eq{Syn:VerticalVelocity} we have used that the vertical velocity in a QG flow is by one order in the Rossby number smaller than expected on the basis of the aspect ratio scaling because the leading-order horizontal divergence vanishes under geostrophic balance \citep[see, e.g.,][]{Pedlosky1992}.

To streamline the notation, we use the following abbreviations in the rest of this section
\begin{equation}\label{eq:qg_quantities}
\left(
\vp, \vpprime, \w, \pi, \Thetaprime \right)
\equiv
\left(
\vp\zero, \vp\order{1}, \w\order{1}, \pi\two / \Gamma, \Theta\two \right)\, .
\end{equation}

Using \eq{eq:qg_quantities} at the leading order in $\eps$, the momentum equations, the mass balance, and the potential temperature transport equation yield
\begin{subequations}\label{GeostrophicEqs}

\noindent
{\slshape Geostrophic Balance}
\vspace{-0.4cm}
\begin{equation}
  \label{GeostrophicBalance}
f_0 \, \vk \times \vp +
\gradhor{\pi} = 0\, ,
\end{equation}

\noindent
{\slshape Hydrostatic Balance}
\vspace{-0.4cm}
\begin{equation}
  \label{HydrostaticBalance}
  \pp{\pi}{z} = \Thetaprime\, ,
\end{equation}

\noindent
{\slshape Anelastic Constraint}
\vspace{-0.4cm}
\begin{equation}
  \label{AnelasticConstraint}
  \rho_0\, \gradhor{} \cdot \vpprime + \pp{}{z}\Big (\rho_0 \, w \Big) = 0\, ,
\end{equation}

\noindent
{\slshape Potential Temperature Transport}
\vspace{-0.4cm}
\begin{equation}
  \label{PotentialTemperatureTransport}
  \pp{\Thetaprime}{t} + \vp \cdot \gradhor{\Thetaprime} + w \dd{\Theta_1}{z} = Q^{\rm QG}_{\Theta} \, ,
\end{equation}
\end{subequations}
where we have assigned the superscript $\rm QG$ to the source term in \eqref{PotentialTemperatureTransport} to indicate that it represents a source term with asymptotic scaling compatible with the QG regime (see the discussion in section~\ref{sec:Intro},\ref{ssec:Motivation}).

\noindent
For later reference, we conclude from \eq{GeostrophicBalance} that
\begin{equation}
  \label{ResultVZeroDivFreeAndGeoWind}
  \gradhor\cdot \vp = 0            
  \qquad\text{and}\qquad
  \vp = \frac{1}{f_0} \vk \times \gradhor\pi\, .
\end{equation}
An additional equation for the divergence of the perturbation velocity, $\gradhor{} \cdot \vpprime$, appearing in \eq{AnelasticConstraint} is obtained from the curl of the next-order horizontal momentum equation,

\noindent
{\slshape Vorticity Transport}
\begin{equation}
  \label{VorticityTransport}
  \bigg(\pp{}{t}+ \vp\cdot\gradhor{}\bigg) \big( \zeta + \beta y
  \big) + f_0\gradhor{}\cdot\vpprime = 0\, ,
\end{equation}
with the relative vertical vorticity
\begin{equation}
\label{eq:vorticity_definition}
  \zeta = \vk\cdot\left(\gradhor\times\vp\right)\, .
\end{equation}

Eqs.\ \eq{GeostrophicEqs}--\eq{eq:vorticity_definition} constitute the QG model for the unknowns $\Big(\vp, w,  \pi, \Thetaprime, \left[\gradhor{}\cdot\vpprime\right]\Big)(t,\vx,z)$ defined in \eq{eq:qg_quantities} given the background state through $\rho_0(z)$ and $\left(\dThetaonedz\right)(z)$. The essence of the system is revealed through its classical formulation involving an advection equation
\begin{equation}\label{eq:qg_pv_advection}
\left(\pp{}{t}+ \vp\cdot\gradhor{}\right) q = \frac{f_0}{\rho_0}\pp{}{z}\left(\frac{\rho_0 \, \QTheta^{\rm QG}}{d\Theta_1/dz}\right)
\end{equation}
for the QG potential vorticity
\begin{equation}\label{eq:qg_pv}
q = \zeta + \beta \, y + \frac{{f_0}}{\rho_0} \pp{}{z} \left(\frac{\rho_0\, \Thetaprime}{d\Theta_1/dz}\right)
\, .
\end{equation}
To verify these equations, one eliminates $\left[\gradhor{}\cdot\vpprime\right]$ from \eq{VorticityTransport} using \eq{AnelasticConstraint}, and then eliminates $w$ from the remaining equation using \eq{PotentialTemperatureTransport}, and finally noticing that $\pp{\vp}{z}\cdot\gradhor \Thetaprime = 0$ follows straight from \eq{GeostrophicBalance} and \eq{HydrostaticBalance}. Given the potential vorticity field $q$ at any time, the Exner pressure field can be recovered by utilizing the hydrostatic balance from \eq{HydrostaticBalance} and the divergence of the geostrophic balance \eq{GeostrophicBalance}, which gives
\begin{equation}
\zeta = \frac{1}{f_0}\gradhor^2 \pi  \,,
\end{equation}
and inserting into \eq{eq:qg_pv} to obtain
\begin{equation}\label{eq:qg_Poisson}
\gradhor^2 \pi + \frac{{f_0^2}}{\rho_0} \pp{}{z} \left(\frac{\rho_0}{\dzThetaone}\pp{\pi}{z}\right)
= f_0 (q-\beta y)\,.
\end{equation}
We note in passing that QG theory obviously relies on a stable background stratification with $\Theta_1' = d\Theta_1/dz \not= 0$ everywhere, as seen from the second term on the left of \eq{eq:qg_Poisson} which involves this quantity in the denominator.


\subsection{The Ekman layer} 
\label{ssec:EkmanLayer}

The flow in the QG layer has been modelled essentially as frictionless. Near the ground, however, friction is responsible to guarantee compliance with a no-slip or related surface boundary conditions. Ekman layer theory \citep[see, e.g.,][]{Pedlosky1992} describes this influence of friction. In the context of QG flows, and under the assumption of a constant turbulent friction coefficient, Pedlosky shows that there is a vertical massflux out of or into the friction layer that is proportional to the vertical vorticity near the ground. By asymptotic matching, this generates an effective bottom boundary condition for the surface vertical velocity within the QG theory, i.e., 

\medskip\noindent
{\slshape within QG theory:}
\begin{equation}
\eps \, w\order{1}_{\text{QG}}(t,\vx, 0) = \frac{\sqrt{E_V}}{2} \, \zeta\order{0}_{\text{QG}}(t,\vx,0)
\end{equation}
where $\zeta = v_x - u_y$ is the vertical vorticity, and 
\begin{equation}
E_V = \frac{2 A_V}{f_0 \hsc^2}
\end{equation}
is the vertical Ekman number based on the pressure scale height $\hsc$ and the turbulent friction coefficient $A_V$. The usual assumption in the coupling of an Ekman layer with QG flow is that 
\begin{equation}
E^*_V = \frac{E_V}{\eps^2} = \bigo{1}
\qquad (\eps \to 0) \,,
\end{equation}
so that there is a vertical flow at the top of the Ekman layer that imposes a nontrivial bottom boundary condition for $w\order{1}$ within the QG theory.   


\section{Geostrophically balanced \thelayer}
\label{sec:WeatherLayer}

In the lower few kilometers, owing to the influence of moist and other strong diabatic processes, we expect the potential temperature stratification to vary on a shorter vertical characteristic scale than it does in the bulk of the troposphere. Thus we introduce an intermediate layer of thickness $\bigo{\sqrt{\eps}\hsc}$ resolved by a stretched vertical coordinate
\begin{equation}
\eta = \frac{z}{\delta}
\qquad\text{where}\qquad
\delta \equiv \sqrt{\eps}\,.
\end{equation}
Within this layer, the potential temperature expands as 
\begin{equation}\label{eq:WLThetaExpansion}
\Theta = 1 + \delta^2 \Theta_1(0) + \delta^3 \Theta\order{3/2} + \littleo{\delta^3}\,,
\end{equation}
where $\Theta\order{3/2}$ is a time dependent, three-dimensional field, i.e., 
\begin{equation}\label{eq:WLSignature}
\Theta\order{3/2} = \Theta\order{3/2}(t,\vx,\eta)\,.
\end{equation}
In this section we label the asymptotic expansion functions, such as $\Theta\order{3/2}$, by integer multiples of $1/2$ to indicate that we are expanding in terms of powers of $\delta = \eps^{1/2}$ and to distinguish the expansion functions introduced here from those utilized in the QG expansions of the previous section. Thus, e.g., $\phi^{(1)}$ would be a variable from QG theory whereas $\phi^{(2/2)}$ would be the corresponding expansion term for the same physical quantity at the same order in $\eps$ in the diabatic layer.

For a layer of this thickness, and provided the horizontal velocity magnitude remains comparable to that in the bulk troposphere, turbulent friction will play a role only at higher orders. Indeed a rescaling of the vertical coordinate by $\eps$ instead of by $\sqrt{\eps}$ is needed in the Ekman layer to lift turbulent friction to a leading-order effect (see the discussion in section~\ref{sec:QGAndEkman}\ref{ssec:EkmanLayer} above). As a consequence, the \thelayer\ physics differs from that seen in the bulk of the troposphere mainly by the scaling of potential temperature perturbations as expressed in \eq{eq:WLThetaExpansion}. 

The particular asymptotic ansatz in \eq{eq:WLThetaExpansion} guarantees that the total variation of potential temperature across the layer is comparable in magnitude to that seen also in the QG regime. In fact, consider some  height $z = \sqrt{\eps} \, \eta$ for a fixed value of the stretched coordinate $\eta$. The QG-layer expansion of potential temperature from \eq{Syn:ThetaExpansion} yields, via Taylor expansion, $\Theta = 1 + \eps\Theta_1(0) + \eps^{\frac{3}{2}}\eta\, \Theta_1'(0) + \text{h.o.t.}$ (see also \eq{eq:ThetaExpansionNearGround} in the appendix), where the prime notation represents the derivative with respect to the unscaled vertical coordinate, $z$, as before. Thus, the deviation of potential temperature from its value at the ground is $\bigo{\eps^{\frac{3}{2}}}$, i.e., comparable to the deviations allowed for in diabatic layer expansion in \eq{eq:WLThetaExpansion}. The key new aspect of the latter is, however, that flow-induced spatio-temporal variations of potential temperature are no longer small perturbations away from a given, approximately linear, background stratification but appear directly in the leading order variability, since the first two terms in \eq{eq:WLThetaExpansion} are constants. See also the right panel of fig.~\ref{fig:QG_vs_QGWL} for illustration.

The remainder of this section summarizes the derivation of the \thelayer\ dynamics, which differs from QG dynamics and reveals a decisive influence of diabatic and moisture effects on the lower troposphere. 


\subsection{Expansion scheme for the \thelayer} 
\label{ssec:WLExpansions}

The expansion scheme in the new layer reads
\begin{subequations}\label{eq:WLSynopticExpansions}
\begin{alignat}{7}
  \label{Syn:WLPressureExpansion}
  &\pi
  && = 1
  && + \sqeps\, \pihalf 
  && + \deltaorder{3}\, \pithreehalf 
  && + \deltaorder{4}\, \pi\order{4/2} 
  && + \deltaorder{5}\, \pi\order{5/2} 
  && + \littleo{\deltaorder{5}}
  \\
  \label{Syn:WLDensityExpansion}
  & \rhonull
  && = 1
  && + \sqeps\, \rhohalf
  && + \deltaorder{2}\, \rho_{2/2} 
  && + \deltaorder{3}\, \rho\order{3/2} 
  && + \littleo{\deltaorder{3}}
\end{alignat}
\begin{alignat}{6}
\label{Syn:WLHorizontalVelocity}
  & \vp
  && = \phantom{\deltaorder{2} } \vp\order{0/2}
  && + \delta \vp\order{1/2}
  && + \littleo{1}
  \\
\label{Syn:WLVerticalVelocity}
  & w
  && = \deltaorder{2} w\order{2/2} 
  && + \littleo{\deltaorder{2}}   
\end{alignat}
\end{subequations}
where $\left(\vp, w, \pi, \rho, \Theta\right)\order{i/2}(t,\vx,\eta)$ have the same dependencies as those listed for $\Theta\order{3/2}$ in \eq{eq:WLSignature}.

Note that in \eq{eq:WLSynopticExpansions} we dropped some intermediate order terms from the expansions to further streamline the notation and derivations below. These expansions lead to a closed system of leading order equations so that they are consistent with the scaling regime considered. None of the omissions is ad hoc, however. To the contrary, the absence of the omitted intermediate level perturbations can be justified by first including these terms and then demonstrating that they must vanish for a consistent expansion (not shown).


\subsection{\Thelayer\ governing equations}
\label{ssec:WLTheory}

Inserting \eq{eq:WLThetaExpansion}, \eq{eq:WLSynopticExpansions} into the dimensionless governing equations from \eq{eq:NondimensionalCompressibleFlowEquations} keeping in mind the definition of $\delta = \sqrt{\eps}$ we obtain the following leading order equations 

\begin{subequations}\label{eq:WLGeostrophicEqs}
\noindent
{\slshape Geostrophic Balance}
\vspace{-0.4cm}
\begin{equation}
  \label{eq:WLGeostrophicBalance}
f_0 \, \vk \times \vp\order{0/2} +
\frac{1}{\Gamma}\gradhor{\pi\order{4/2}} = 0\, ,
\end{equation}

\noindent
{\slshape Hydrostatic Balance}
\vspace{-0.4cm}
\begin{equation}
  \label{eq:WLHydrostaticBalance}
  \frac{1}{\Gamma}\pp{\pi\order{4/2}}{\eta} = \Theta\order{3/2}\, ,
\end{equation}

\noindent
{\slshape Mass Balance}
\vspace{-0.4cm}
\begin{equation}
  \label{eq:WLAnelasticConstraint}
  \gradhor{} \cdot \vp\order{0/2} = 0\, ,
\end{equation}

\noindent
{\slshape Potential Temperature Transport}
\vspace{-0.4cm}
\begin{equation}
  \label{eq:WLPotentialTemperatureTransport}
  \left(\pp{}{t} + \vp\order{0/2} \cdot \gradhor{} \right) \Theta\order{3/2} 
  = \left(Q_\Theta + S_\Theta\right)\order{3/2} \, .
\end{equation}
\end{subequations}
These equations must be supplemented by initial conditions for $\Theta\order{3/2}$, by a closure for the source terms on the right hand side of \eq{eq:WLPotentialTemperatureTransport}, and by the top-of-the-boundary layer matching conditions. The latter will be discussed in subsection~\ref{ssec:DLBulkMatching} below. 

Information regarding the leading order vertical velocity in the \thelayer\ is obtained from the next order expansions. The horizontal momentum balance at the next order is
\begin{equation}
  \label{eq:WLGeostrophicBalanceNext}
f_0 \, \vk \times \vp\order{1/2} +
\frac{1}{\Gamma}\gradhor{\pi\order{5/2}} = 0\, ,
\end{equation}
and this implies 
\begin{equation}
\gradhor\cdot\vp\order{1/2} = 0\,.
\end{equation}
Now, since the next order mass balance is
\begin{equation}
  \label{eq:WLAnelasticConstraintNext}
  \gradhor\cdot\vp\order{1/2} + \pp{w\order{2/2}}{\eta} = 0
\end{equation}
we conclude that
\begin{equation}
  \label{eq:WLWNextStructure}
  w\order{2/2}(t,\vx,\eta) \equiv \widehat w\order{2/2}(t,\vx)
\end{equation}
is independent of $\eta$. 


\subsection{Matching to the QG flow} 
\label{ssec:DLBulkMatching}

To obtain a physically consistent smooth transition between the solution in the \thelayer\ and the QG flow in the bulk of the troposphere, the two solutions have to agree asymptotically (in the limit $\eps\to 0$) in an overlap region.

The horizontal velocity $\vp\order{0/2}$ is entirely determined by the Exner pressure perturbation field $\pi\order{4/2}$ through the geostrophic balance in \eq{eq:WLGeostrophicBalance}. Since this equation is the exact equivalent of the corresponding geostrophic balance equation for the bulk flow from \eq{GeostrophicBalance}, the horizontal velocities in the QG and DL regions will automatically match in an overlap region between the two layers if the pressure perturbations $\pi\order{2}$ and $\pi\order{4/2}$, respectively, do so.

The vertical velocity is constant in the vertical direction in the DL as shown in~\eq{eq:WLWNextStructure}. As a consequence, matching vertical velocities between the QG and DL flows implies that 
\begin{equation}\label{eq:DLVerticalVelocity}
w\order{2/2}(t,\vx,\eta) \equiv \widehat w\order{2/2}(t,\vx) = w\order{1}(t,\vx,0)\,.
\end{equation}

We derive the required large $\eta$ behavior of $\Theta\order{3/2}$ and $\pi\order{4/2}$ by van Dyke's approach \citep[see][]{vanDyke1975,Eckhaus1979}. Thus we expand the QG solutions for $z \ll 1$, replace $z = \delta \eta$ in the resulting expressions, and then compare like powers of $\delta$ with the DL solutions. For the potential temperature this yields (see \eq{Syn:ThetaExpansion} and \eq{eq:WLThetaExpansion})
\begin{equation}
1 + \delta^2 \Thetaone(0) + \delta^3 \Thetaone'(0) \eta 
= 1 + \delta^2 \Thetaone(0) + \delta^3 \Theta\order{3/2}(t,\vx,\eta) + \text{h.o.t}\,.
\end{equation}
That is, we require
\begin{equation}\label{eq:ThetaThreeHalvesAsymptotics}
\frac{\Theta\order{3/2}(t,\vx,\eta)}{\eta} - \Thetaone'(0)  \to 0 
\qquad (\eta \to \infty) \,,
\end{equation}
or equivalently $\delta \to 0$ at fixed $z$. Considering the evolution equation for $\Theta\order{3/2}$ in \eq{eq:WLPotentialTemperatureTransport}, this condition will be met if it is met initially at $t=0$ and if the driving terms $\left(Q_\Theta + S_\Theta\right)\order{3/2}$ vanish sufficiently rapidly for large $\eta$. The latter condition must be satisfied if QG theory is to be applicable in the bulk of the troposphere, i.e., above the bottom boundary layers, in the first place. 

As regards matching of the Exner pressure, we utilize the fact that in both the QG and the DL regions we have hydrostatic balance. Accordingly, in the QG layer the Taylor expansion for $z \ll 1$ with the replacement $z = \delta \eta$ reads (see the appendix)
\begin{alignat}{2}
\left.\pi(t,x,z;\eps)\right|_{\text{QG}} 
  & = 1 - \delta \Gamma \eta 
     + \delta^3 \Gamma \Theta_1(0) \eta
   \label{eq:QGExnerNearGround}\\
  & + \delta^4 \left(\pi\order{2}(t,\vx,0) + \Gamma \Theta_1'(0) \frac{\eta^2}{2}
                \right)
     + \text{h.o.t.}\,.
     \nonumber
\end{alignat}
The Exner pressure representation in the \thelayer\ is
\begin{alignat}{2}
\left.\pi(t,x,z;\eps)\right|_{\text{DL}} 
  & = 1 - \delta \Gamma \eta
      + \delta^3 \Gamma \Theta_1(0) \eta
    \label{eq:DLExnerLargeEta}\\
  & + \delta^4 \left(\pi\order{4/2}(t,\vx,0)  
                       + \Gamma \int\limits_{0}^{\eta} \Theta\order{3/2}(t,\vx,\eta')\, d\eta'
                 \right) 
    \nonumber\\
  &  + \text{h.o.t.}\,.
     \nonumber
\end{alignat}
Provided that $\Theta\order{3/2}$ approaches its asymptotic behavior from \eq{eq:ThetaThreeHalvesAsymptotics} sufficiently rapidly, we conclude that 
\begin{equation}
\pi\order{4/2}(t,\vx,0) 
= \pi\order{2}(t,\vx,0) + \Delta\pi\order{4/2}_{\text{DL}}(t,\vx)\,,
\end{equation}
where the effective pressure change across the DL reads
\begin{equation}\label{eq:DLExnerChange}
\Delta\pi\order{4/2}_{\text{DL}}(t,\vx) 
= - \Gamma\int\limits_{0}^{\infty} \left[\Theta\order{3/2}(t,\vx,\zeta) - \Thetaone'(0) \zeta \right] d\zeta\,.
\end{equation}
Importantly, this shows that the dynamically relevant pressure field changes substantially across the \thelayer\ as discussed in section~\ref{sec:Essence}.


\subsection{Matching to the Ekman layer} 
\label{ssec:DLEkmanMatching}

Bottom boundary conditions for the \thelayer\ are provided by matching to the Ekman layer from section~\ref{sec:QGAndEkman}\ref{ssec:EkmanLayer}. The latter remains entirely unchanged, except that the vorticity at the bottom of the QG flow in section~\ref{sec:QGAndEkman}\ref{ssec:EkmanLayer} is to be replaced by that at the bottom of the DL, i.e, we have

\medskip\noindent
{\slshape with QG-DL-Ekman theory:}
\begin{equation}
w\order{1}_{\text{DL}}(t,\vx, 0) 
= \frac{\sqrt{E^*_V}}{2} \, \zeta\order{0}_{\text{DL}}(t,\vx,0)
= w\order{1}_{\text{QG}}(t,\vx, 0)\,.
\end{equation}
The second equality results from the fact that throughout the \thelayer\ the vertical velocity $w\order{1}$ is independent of the vertical coordinate, $\eta$, according to \eq{eq:DLVerticalVelocity}, so that $w\order{1}_{\text{QG}}(t,\vx, 0) = w\order{1}_{\text{DL}}(t,\vx, 0)$ as announced in section~\ref{sec:Essence}.


\section{Summary and Discussion}
\label{sec:SummaryAndDiscussion}

In this paper we have extended the classical quasi-geostrophic / Ekman layer theory \citep[see][]{Pedlosky1992} by including a low-altitude ``\thelayer'' (DL) of intermediate thickness. Within this layer, dynamically evolving potential temperature variations arise not as small deviations from a stable background stratification, but are instead comparable to the latter. Moreover, the synoptic-scale horizontal mean stratification in the DL is not restricted to being stable. The situation is sketched in fig.~\ref{fig:QG_vs_QGWL}.

This latter result is important in the light of a proper interpretation of the role of the DL for quasi-geostrophic theory: According to the classical theory, QG solutions in the bulk of the troposphere depend crucially on the vertical velocity near the ground as generated by orography or by the outflow from the Ekman friction layer of height $\bigo{\eps\hsc}$, see fig.~\ref{fig:Feedbacks}. In the present theory, this vertical velocity equals that at the bottom of the DL of height $\bigo{\sqrt{\eps}\hsc}$, i.e., $w\order{2/2}(t,\vx,0)$ because, according to \eq{eq:WLAnelasticConstraintNext}, $w\order{2/2}$ is homogeneous in the vertical coordinate $\eta$ and is thus mediated without change to the bottom of the bulk tropospheric flow. 

Ekman theory proceeds exactly as known from classical textbooks, so that the Ekman layer outflow vertical velocity can be expressed in terms of the leading order vertical vorticity $\zeta\order{0}$ at the bottom of the next layer, which in the present case is the \thelayer. Since the QG and DL flows are both geostrophic, the vorticity may be expressed in terms of the Exner pressure perturbation field and we arrive at
\begin{equation}
w\order{1}
= \frac{E^*_V}{2f_0} \gradhor^2 \left(\pi\order{2} + \Delta\pi\order{2}_{\text{DL}}\right)\,.
\end{equation}
Here $\Delta\pi\order{2}_{\text{DL}}$ is the (vertical) change of the dynamically relevant Exner pressure variation across the DL. Thus we conclude that diabatic and moist processes in this layer influence the QG flow at leading order by contributing to pressure and thus vorticity variations on top of the Ekman layer which then, in turn, determine the vertical velocity also at the lower boundary of the QG layer.

Corroboration of the existence of a dynamically relevant layer of $~\unit{3}{\kilo\meter}$ height in the lower troposphere within which the energetically dominant part of moist processes take place is found in in the literature. For instance, \citet{WoodBretherton2006} discuss how lower troposphere statistic stability plays a strong role in stratiform low cloud development, and \citet{YueEtAl2011} argue that strong variations in static stability are commonly seen in the ITCZ and other stormy regions of the globe.  

As an outlook to future work we mention that in the present paper we have adopted the dry air QG theory for the description of the bulk of the troposphere. A promising extension of the present work would be to adopt the recent moist QG model developed by \citet{SmithStechmann2017} \citep[see also][]{MarsicoEtAl2019,WetzelEtAl2019} and to include the \thelayer. Moist processes in the DL will likely have a multiscale character, and a their thorough analysis may proceed along the lines of \citet{MajdaKhouider2002,KhouiderMajda2006,OwinohEtAl2011,HittmeirKlein2018} allowing for multiple flow features (shallow and deep for example) and, more importantly, cloud types to evolve and interact with each other. Another current development concerns explicit frontal solutions of the new three-layer model that would include known QG fronts in the bulk troposphere and a matching flow structure in the \thelayer. This would provide idealized mutual tests for the theory on the one hand, and for full-fledged weather forecast models on the other.

\acknowledgments
L.S., S.P., and R.K.'s work has been funded by Deutsche Forschungsgemeinschaft (DFG) through grant CRC~1114 ``Scaling Cascades in Complex Systems'', Project Number 235221301, Project (C06) ``Multi-scale structure of atmospheric vortices''. This research of BK is partially supported by a Natural Sciences and Engineering Research Council of Canada Discovery grant.

%
%
\datastatement
The paper presents theoretical work and all derivations should be described in sufficient detail. No further data are required to reproduce the findings.

%
\appendix[A]



\appendixtitle{Matching of pressures between the QG and intermediate layers}
\label{Asec:MatchingQGWL}

As regards matching of the Exner pressure we utilize that in both the QG and intermediate layers we have hydrostatic balance. Accordingly, in the QG layer the Taylor expansion near $z=0$ with the replacement $z = \delta \eta$ reads
\begin{alignat}{2}
&\pi(t,x,z) 
  \\
&= 1 + \delta^4 \pi\order{2}(t,\vx,0) - \int_0^z \frac{\Gamma}{\Theta(t,\vx,z')}\, dz'
   \nonumber\\
&= 1 + \delta^4 \pi\order{2}(t,\vx,0) 
     - \Gamma \left(z - \delta^2 \left[\Theta_1(0)  z 
                                     - \Theta_1'(0) \frac{z^2}{2} 
                                     + ...
                                 \right]
              \right)
   \nonumber\\
&= 1 + \delta^4 \pi\order{2}(t,\vx,0) 
     - \Gamma \left(\delta \eta - \delta^3 \Theta_1(0) \eta 
                                - \delta^4 \Theta_1'(0) \frac{\eta^2}{2} 
                                + ...
              \right)
   \nonumber\\
&= 1 - \delta \Gamma \eta 
     + \delta^3 \Gamma \Theta_1(0) \eta
     + \delta^4 \left(\pi\order{2}(t,\vx,0) + \Theta_1'(0) \frac{\eta^2}{2}
                \right)
     + ...
  \nonumber
\end{alignat}
where we have used that
\begin{alignat}{2}
\frac{1}{\Theta} 
& = \frac{1}{1 + \delta^2 \Theta_1 + \delta^4 (\Theta\order{2}+ \Theta_2 ) + ...} 
  \\
&= 1 - \delta^2 \Theta_1 - \delta^4 \left(\Theta\order{2} + \Theta_2 - \frac{1}{2} \Theta_1^2 \right) 
 + ...
 \nonumber
\end{alignat}
as well as the Taylor expansion of $\Theta_1$ for $z \ll 1$,
\begin{equation}\label{eq:ThetaExpansionNearGround}
\Theta_1(z) = \Theta_1(0) + \Theta_1'(0) z + \Theta_1''(0) \frac{z^2}{2} + ...\,.
\end{equation}
%

%
\bibliographystyle{ametsoc2014}
\bibliography{./References}

\end{document}